\documentstyle[amsfonts,12pt]{article}
\newcommand{\be}{\begin{equation}}\newcommand{\ee}{\end{equation}}
\newcommand{\bea}{\begin{eqnarray}}\newcommand{\eea}{\end{eqnarray}}
\newcommand{\nn}{\nonumber \\}\newcommand{\p}[1]{(\ref{#1})}

\newcounter{muni}

\begin{document}
\begin{titlepage}
\begin{flushright}
LPTHE 98-43 \\
JINR E2-98-248 \\
hep-th/9809108 \\
September 1998
\end{flushright}
\vskip 1.0truecm
\begin{center}
{\large \bf Quaternionic Taub-NUT from the harmonic space approach}
\end{center}
\vskip 1.0truecm
\centerline{\bf Evgeny Ivanov${}^{(a)}$, Galliano Valent${}^{(b)}$}
\vskip 1.0truecm
\centerline{${}^{(a)}$\it Bogoliubov Laboratory of Theoretical 
Physics, JINR,}
\centerline{\it Dubna, 141 980 Moscow region, Russia}
\centerline{\tt eivanov@thsun1.jinr.ru}
\vskip5mm
\centerline{${}^{(b)}$ \it 
Laboratoire de Physique Th\'eorique et des Hautes Energies,}
\centerline{\it Unit\'e associ\'ee au CNRS URA 280,~Universit\'e Paris 7}
\centerline{\it 2 Place Jussieu, 75251 Paris Cedex 05, France}
\centerline{\tt valent@lpthe.jussieu.fr}
\vskip 1.0truecm  \nopagebreak

\begin{abstract}
We use the harmonic space technique to construct explicitly a quaternionic 
extension of the Taub-NUT metric. It depends on two parameters, the first 
being the 
Taub-NUT `mass' and the second one the cosmological constant.
\end{abstract} 
\end{titlepage}

\noindent {\bf 1.} An efficient way to explicitly construct 
hyper-K\"ahler and quaternionic-K\"ahler metrics is provided by the harmonic 
(super)space method \cite{gikos} - \cite{gio}. 

It was firstly introduced in the context 
of $N=2$ supersymmetry \cite{gikos}. 
The basic idea was to extend the standard $N=2$
superspace by a set of internal (`harmonic') variables 
$u^{\pm i}, u^{+i}u^-_i = 1$, 
parametrizing 
the automorphism group $SU(2)$ of $N=2$ superalgebra. 
It was shown in \cite{gikos} 
that all $N=2$ theories admit a manifestly supersymmetric off-shell 
description in terms of unconstrained superfields given on an 
analytic subspace of the $N=2$ harmonic superspace, harmonic {\it analytic} 
superfields. 

It was soon realized that the harmonics are also relevant to 
some purely bosonic geometric problems. 
As is shown in \cite{HK}, the constraints defining the hyper-K\"ahler (HK) 
geometry 
can be given an interpretation 
of the 
integrability conditions for the existence of analytic fields in a $SU(2)$ 
harmonic extension of 
the original $4n$-dimensional HK manifold $\{x^{i\mu}\}, 
(i=1,2; \mu = 1, ... 2n) $.  
This time, the $SU(2)$ to   
be `harmonized' is an extra $SU(2)$ rotating three complex structures of 
the HK manifold. The analytic 
subspace is spanned by the harmonic variables $u^{\pm i}$ and half 
of the initial $x$-coordinates, $x^{+\mu}$.
The constraints of HK geometry can be solved via an unconstrained 
analytic HK potential ${\cal L}^{+4}(x^{+\mu}, u^{\pm i})$. It encodes 
(at least, locally) all the information about the associated metric. 
Remarkably, it allows one to {\it explicitly} construct 
the HK metrics by simple rules \cite{HK}.

In \cite{gio}, a generalization of this approach to the  
quaternionic-K\"ahler (QK) manifolds was given. 
These manifolds generalize the HK ones in that the extra $SU(2)$ 
which transforms complex structures becomes an 
essential part of the holonomy group. 
It was shown in \cite{gio}, that 
the QK geometry constraints can be also solved 
in terms of
some unconstrained potential ${\cal L}^{+4}$ living on the analytic 
subspace 
parametrized by $SU(2)$ harmonics and half of the original  
coordinates. 
The specificity of the QK case is the presence of a
non-zero 
constant $Sp(1)$ curvature on all steps of the way from  
${\cal L}^{+4}$ to the related metric. 
It is interesting to consider some examples in order to see 
in detail how the 
machinery proposed in \cite{gio} works. Only the simplest case of the 
homogeneous QK manifold  $Sp(n+1)/Sp(1)\times Sp(n)$ (corresponding to 
${\cal L}^{+4} = 0$) was considered in ref. \cite{gio}. 

The aim of this paper is to demonstrate the power of the harmonic 
geometric approach on the example of less trivial QK metric, 
a quaternionic 
generalization of the well-known four-dimensional Taub-NUT (TN)
metric \cite{egh}. 
Like in the HK case \cite{cmp}, the computations are greatly 
simplified due to the $U(1)$ isometry of the quaternionic  
TN metric. 
The metric depends on two 
parameters, the TN `mass' parameter and the constant 
$SU(2)$ curvature parameter which 
can be interpreted as the inverse `radius' of the corresponding `flat' QK 
background $\sim Sp(2)/Sp(1)\times Sp(1)$. We perform the 
identification of the metric with those known in literature and 
consider its few important particular limits. 

\vspace{0.4cm}
\noindent {\bf 2.} We first recall some salient features of the 
construction of \cite{gio}. 
One starts with a $4n$-dimensional Riemann manifold parametrized by 
local coordinates $\{x^{\mu m}\}, \mu = 1,2, ..., 2n;$ $m = 1,2$, and 
uses a vielbein formalism. 
The QK geometry can be defined as a restriction of the general Riemannian 
geometry in $4n$-dimensions, such that the holonomy group of 
the corresponding manifold is a subgroup of $Sp(1)\times Sp(n)$
\footnote{For 
the $4$-dimensional case this definition has to be replaced by 
the requirement that the totally symmetric 
part of the $Sp(1)$ component of the curvature tensor lifted to the 
tangent space is vanishing.}. 
Thus one can choose the tangent space group from the very beginning to be 
$Sp(1)\times Sp(n)$ and define the QK geometry via appropriate 
restrictions on the curvature tensor lifted to the tangent space 
(taking into account that the holonomy group is generated 
by this tensor). As explained in \cite{gio},  
for the QK manifold of generic dimension the defining constraints can 
be written as a restriction on the commutator
of two covariant derivatives
\be 
\left[ {\cal D}_{\alpha (i}, {\cal D}_{\beta k)} \right] =
 -2 \Omega_{\alpha\beta}
R \Gamma_{(ik)}~. \label{constr}
\ee         
Here 
\be \label{covD}
{\cal D}_{\alpha i}  = e^{\mu m}_{\alpha i}(x) {\cal D}_{\mu m}  
= e^{\mu m}_{\alpha i}(x)\;\frac{\partial}{\partial x^{\mu m}} 
+ [Sp(1)\times Sp(n)-\mbox{connections}] ~,
\ee
$e^{\mu m}_{\alpha i}(x)$ being the
$4n \times 4n$ vielbein with the indices $\alpha = 1,2, ... 2n$ 
and $i=1,2$ 
rotated, respectively, by the tangent local $Sp(n)$ and $Sp(1)$ groups, 
$\Omega_{\alpha\beta}$ is the $Sp(n)$-invariant skew-symmetric tensor 
serving to raise and lower the $Sp(n)$ indices 
($\Omega_{\alpha\beta}\Omega^{\beta\gamma} = \delta^\gamma_\alpha $), 
$\Gamma_{(ik)}$ are the $Sp(1)$ generators, and  $R$ is a constant, 
remnant of the $Sp(1)$ component of the Riemann tensor (its constancy is 
a consequence of the QK geometry constraint and Bianchi identities). 
The scalar curvature coincides with $R$ up to a positive 
numerical coefficient, so the cases $R>0$ and $R<0$ correspond to compact 
and non-compact manifolds, respectively. 
In the limit $R=0$ eq. \p{constr} is reduced to the constraint defining 
the HK geometry \cite{HK}, in accord with the interpretation of HK 
manifolds as a degenerate subclass of the QK ones. 
 
Like in the HK case \cite{HK}, in order to explicitly figure out which 
kind of restrictions is imposed by (1) on the vielbein 
$e^{\mu m}_{\alpha i}(x)$ and, hence, on the metric 
\be  \label{metr}
g^{\mu m\; \nu s} = e^{\mu m}_{\alpha i}\; e^{\nu s \;\alpha i}~, \quad 
g_{\mu m \; \nu s} = e_{\mu m \;\alpha i}\; e_{\nu s}^{\alpha i}~,
\ee
one should solve the constraints \p{constr} 
by regarding them as integrability conditions along some complex 
directions in a harmonic extension of the original manifold. 

Due to the non-vanishing r.h.s. in (1), the road to such an interpretation 
in the QK case is more tricky. Modulo these peculiarities, the 
basic step still consists in extending $\{x^{\mu m} \}$ by a set 
of some harmonic variables, $\{x^{\mu m} \} \rightarrow 
\{x^{\mu m}, w^\pm_i \}, \;w^{+\;i}w^-_i = 1$. 
Then, following the general 
strategy \cite{HK}, \cite{gio}, one passes to a new (`analytic') basis in 
$\{x^{\mu m}, w^\pm_i \}$
\bea
&& \{x^{\mu m}, w^\pm_i \}  \Rightarrow 
\{x^{+\mu}_A, x^{-\mu}_A, w^{\pm\;i}_A \}
\label{passCB} \\
&& x^{\pm \mu}_A = x^{\mu\; i}w^{\pm}_i + v^{\pm \mu}(x,w)~, \quad
w^{+\;i}_A = w^{+\;i} -R v^{++}(x,w)w^{-\;i}~, \quad w^{-\;i}_A = 
w^{-\;i}~, 
\label{xbridge} 
\eea
where the `bridges' $v^{\pm \mu}(x,w), v^{++}(x,w)$ are chosen so as to 
make the $w^+$-projection of ${\cal D}_{\alpha i}$ in this basis to be 
proportional to the partial derivative with respect to $x^{-\mu}$
\be 
{\cal D}_{\alpha}^+  \sim w^{+i} {\cal D}_{\alpha i} = 
w^{+i}e^{\mu m}_{\alpha i}(x)\partial_{\mu m} +
\dots  = E^\mu_\alpha (x,w) \frac{\partial}{\partial x^{-\mu}_A} =
E^\mu_\alpha (x,w) \partial^+_\mu
\label{Dstruct}
\ee
(simultaneously, one performs an appropriate $Sp(n)$ rotation 
of the tangent space index $\alpha $ by a matrix $Sp(n)$-`bridge'). 
The possibility to reduce ${\cal D}_{\alpha}^+$ to this `short' form 
amounts to the possibility to define {\it analytic} fields living 
on the analytic subspace $\{x^{+\mu}_A, \break w^{\pm\;i}_A \}$. 
The original QK geometry constraints prove to be equivalent to the 
existence of such analytic fields and subspace \cite{gio}.  An essential 
difference of the QK case from the HK case \cite{HK} is the necessity 
to shift the harmonic variables with the new bridge $v^{++}$. 

Besides the opportunity to make ${\cal D}_{\alpha}^+$ short, the 
passing to the harmonic extension of $\{x^{\mu n}\}$ and 
further to the analytic basis and frame (`the $\lambda $-world') 
allows one to exhibit the fundamental 
unconstrained objects of the QK geometry, the QK potential. While 
in the original formulation (`the $\tau $-world') the basic geometric 
objects are the vielbeins $e^{\mu m}_{\alpha k}(x)$ properly constrained 
by eq. \p{constr}, in the analytic basis such objects are 
the harmonic vielbeins covariantizing the derivatives with respect to 
the harmonic variables. In the original basis 
the harmonic derivatives are $D^{\pm\pm} = \partial^{\pm\pm}_w = 
w^{\pm i}\partial/\partial w^{\mp i}\;, \; 
D^0 = \partial^{0}_w = w^{+i} \partial/\partial w^{+i} - 
w^{-i} \partial/\partial w^{- i}\;, \;
[\partial^{++}_w, \partial^{--}_w] = \partial^0\;,$ i.e. they contain 
no partial derivatives with respect to the variables $x^{\mu n}$, because 
the harmonic space $\{x^{\mu m}, w^\pm_i\}$ has the structure of the 
direct product $\{x^{\mu m} \}\otimes \{w^\pm_i \}$. After passing to the 
analytic basis by eqs. \p{xbridge}, the derivatives 
$D^{\pm\pm}$ acquire terms proportional to $\partial^{\pm}_\mu \equiv 
\partial/ \partial x^{\mp \mu}$. Besides, in $D^{++}$ there emerges 
a term proportional to $\partial^{--}_{w_A}$. These 
new terms appear with the appropriate vielbein components 
$H^{+3 \mu}$, $H^{--\pm\mu}$, $H^{+4}$ which are related 
to the bridges as follows 
\bea
&& (\partial^{++}_w + R v^{++}) x^{+\mu}_A = H^{+3\mu}~,  \label{H3} \\
&& (\partial^{++}_w + Rv^{++}) v^{++} = - H^{+4}~, \label{H4} \\
&&  \frac{1}{1-R\partial_w^{--}v^{++}}\;\partial^{--}_w x^{\pm \mu}_A = 
H^{--\pm \mu} ~. \label{Hminplm}  
\eea
Note that $x^{-\mu}$ is determined in terms of $x^{+\mu}$ by 
the equation 
\be
(\partial^{++}_w - R v^{++}) x^{-\mu}_A = x^{+\mu}_A~. \label{x-eq}
\ee 

The original QK geometry constraints require $H^{+3\mu}$, $H^{+4}$ 
to be analytic 
\be
\partial^+_\mu H^{+3\mu} = \partial^+_\mu H^{+4} = 0 \quad 
\Rightarrow H^{+3\mu} = H^{+3\mu}(x^+_A, w_A)~, \; 
H^{+4} = H^{+4}(x^+_A, w_A)~, \label{Hanalyt}
\ee
and express $H^{+3\mu}$ in terms of $H^{+4}$. Basically, 
the analytic harmonic vielbein $H^{+4}$ is just the unconstrained 
QK potential. To be more 
precise, the QK potential ${\cal L}^{+4}$, as it was defined 
in \cite{gio}, 
is related to $H^{+4}$ as (after properly fixing the $\lambda $-world 
gauge freedom)
\be
H^{+4} (x^+_A, w_A)= {\cal L}^{+4}(x^+_A, w_A) + 
x^+_\mu H^{+3\mu}(x^+_A, w_A)~, \qquad x^+_\mu \equiv 
\Omega_{\mu \nu}x^{+\nu}~, \label{QKpot} 
\ee
and 
\bea
H^{+3\nu} = 
{1\over 2}\Omega^{\nu \mu}\hat{\partial}^-_\mu {\cal L}^{+4}~, 
\qquad
\hat{\partial}^-_\mu \equiv \partial^-_\mu + Rx^+_\mu \partial^{--}_A~. 
\label{H3L4}                                        
\eea

It can be shown that the only constraint to be satisfied by ${\cal L}^{+4}$ 
is its analyticity, so this object encodes all the information about 
the relevant QK geometry and metrics, whence its name `QK potential'. 
Choosing one or another explicit ${\cal L}^{+4}$, 
and substituting \p{QKpot}, \p{H3L4} into eqs. \p{H3} - \p{x-eq}, one can 
solve the latter for $x^{\pm \mu}$ and $v^{++}$ as functions of harmonics 
and the $\tau $-world coordinates $x^{\mu m}$. 
Having at hand the explicit form of the variable change \p{xbridge}, 
it remains to find the appropriate expression of the 
$\lambda $-world vielbeins in terms of ${\cal L}^{+4}$ in order to be 
able to restore the $\tau $-world vielbein and hence the QK metric itself. 

Skipping intermediate steps (they can be found in \cite{gio}), 
the non-vanishing components of the $\lambda $-world inverse QK metric 
are given by the following expressions 
\bea
g^{\mu+\;\nu-}_{(\lambda)} = g^{\nu-\;\mu+}_{(\lambda)} = 
\Omega^{\mu\rho}(\partial\hat H)^{-1\;\nu}_{~~\;\;\rho}~, \qquad
g^{\mu-\;\nu-}_{(\lambda)} = 
- 2\;\Omega^{\rho\sigma} (\partial\hat H)^{-1\;\omega}_{~~\;\;\sigma}
(\partial\hat H)^{-1\;(\mu}_{~~\;\;\rho} 
\partial_\omega^+ \hat H^{-3\nu)}~, \label{compmetr}
\eea
where 
\bea
(\partial \hat H)^{\;\mu}_\nu \equiv \partial^+_\nu \hat H^{--+\mu}~, \quad
\hat H^{--\pm \mu} \equiv {1\over 1 - R(x\cdot H)}\;H^{--\pm \mu}~, \quad  
x\cdot H \equiv x^+_\mu H^{--+\mu}~. \label{def1}
\eea
Then the $\tau $-world metric can be obtained via the change of variables 
inverse to \p{passCB}
\be
g^{\mu m \; \nu s} = g^{\omega - \;\sigma -}_{(\lambda)}
\partial^+_\omega x^{\mu m} \partial^+_\sigma x^{\nu s} 
+ g^{\omega + \;\sigma -}_{(\lambda)}\left(
\hat\partial^-_\omega x^{\mu m} \partial^+_\sigma x^{\nu s}
+ \hat\partial^-_\omega x^{\nu m} \partial^+_\sigma x^{\mu s} \right)~.
\label{taumetr}
\ee

In the case of 4-dimensional QK manifolds we will deal with in the sequel
$(\mu, \nu = 1,2 )$ the $\tau $-basis metric \p{taumetr}, after some 
algebra, can be put in the form   
\bea
g^{\mu m \; \nu s} &=& {1\over \mbox{det}(\partial \hat H)}  
\;{1\over [1 -R(x\cdot H)](1- R\partial_w^{--}v^{++})} 
\;G^{\mu m \; \nu s}\;, \label{12} \\
G^{\mu m \; \nu s} &=& \epsilon^{\lambda \rho}\;[\;\partial^{--}_w 
X^{+ \mu m}_\lambda \;X^{+ \nu s}_\rho + 
(\mu m \leftrightarrow \nu s)\;] \;. 
\label{13}
\eea
Here 
\be 
X^{+ \mu m}_\rho \equiv \partial^+_\rho x^{\mu m} \label{defX}
\ee
are solutions of the system of algebraic equations 
\bea 
&& X^{+ \mu m}_\rho \nabla_{\mu m} x^{-\nu} = \partial^+_\rho x^{-\nu} 
= \delta^\nu_\rho, \qquad 
X^{+ \mu m}_\rho \nabla_{\mu m} x^{+\nu} = \partial^+_\rho x^{+\nu} = 0~, 
\label{1} \\
&& \nabla_{\mu m} \equiv \partial_{\mu m} + {R\over 1 - 
R \partial_w^{--} v^{++}}(\partial_{\mu m} v^{++})\partial^{--}_w~.
\eea   

As we see, the problem of calculating the QK metric \p{12}, \p{13} is 
reduced to solving the differential equations \p{H3}, \p{x-eq}, \p{H4} 
which define, by the known ${\cal L}^{+4}(x^{+\mu}, w_A^{\pm i})$, 
$x^{\pm \mu}$ and $v^{++}$ as functions of the $\tau $-basis coordinates 
$x^{\mu m}$ and $w^{\pm i}$. In general, it is a difficult task. 
However, it is simplified for the QK metrics with isometries, like 
in the HK case \cite{cmp}. We will demonstrate this on the example of 
the QK analog of the Taub-NUT metric. 

\vspace{0.4cm}
\noindent {\bf 3.}
The QK counterpart of the TN manifold 
is characterized by the same ${\cal L}^{+4}$ \cite{bgio}
\be 
{\cal L}^{+4} = \left(2i\lambda x^{+} \bar{x}^{+}\right)^2 \equiv 
\left(\phi^{++}\right)^2~. \label{LTN}
\ee
Here we introduced the notation 
\footnote{We adopt the convention 
$\epsilon_{12} = - \epsilon^{12} = 1$. The complex 
conjugation is always understood as a generalized one, i.e. the 
product of the ordinary conjugation and Weyl reflection of harmonics 
\cite{gikos}.}  
\be 
(x^{+1},  x^{+2}) = (x^{+},  - \bar{x}^{+})~, \quad 
\bar x^{+} = \overline{(x^+)}~, \;\; 
\overline{(\bar x^+)} = - x^+~.
\ee
We also assume 
\be
\bar \lambda = \lambda \quad \Rightarrow \quad 
\overline{\phi^{++}} = \phi^{++}~.
\ee
The basic equations \p{H4}, \p{H3}, \p{x-eq} for the given case take 
the form
\bea
&& \partial^{++} v^{++} + R(v^{++})^2 = (\phi^{++})^2~,  \label{veq} \\
&& (\partial^{++}  + Rv^{++}) x^{+} = 
2i\lambda x^{+} \phi^{++}~,\label{xpluseq} \\
&& (\partial^{++}  - Rv^{++}) x^{-} = x^{+}  
\label{xmineq} 
\eea          
(together with their conjugates). These equations are covariant under 
two rigid symmetries preserving the 
analytic subspace $\{ x^+, \bar x^+, w^{\pm i}_A \} $: $U(1)$ Pauli-G\"ursey 
(PG) symmetry 
\be
x^{+\;}{}' = e^{i\alpha}\;x^+~, \qquad  \bar x^{+\;}{}' = 
e^{-i\alpha}\;\bar x^+~,
\label{PG}
\ee
and $SU(2)$ symmetry which uniformly rotates the doublet indices 
of the harmonic variables ($x^{\pm }$ and $v^{++}$ are scalars with 
respect to this $SU(2)$).
They constitute the $U(2)$ isometry group of the QK TN metric.

We will firstly solve eq. \p{veq}. Defining 
\be
v^{++} = \partial^{++}v~, \quad \omega \equiv e^{Rv}~, \quad \hat x^{+} 
\equiv \omega \;x^{+}~, \quad \hat \phi^{++} = 2i \lambda 
\hat x^{+}\bar{\hat x}^{+} = \omega^2 \;\phi^{++}~, \label{defhat}
\ee 
we rewrite \p{veq}, \p{xpluseq} as 
\bea
&& (\partial^{++})^2 \omega = R\; {(\hat\phi^{++})^2 \over \omega^3} 
\label{omeq}~,  \\
&& \partial^{++} \hat x^{+} = 2i\lambda x^+ 
{\hat \phi^{++} \over \omega^2} \equiv 2 i\lambda x^+ \kappa^{++}~. 
\label{hatxeq}
\eea
{}From eq. \p{hatxeq} and the definition of $\hat \phi^{++}$ one 
immediately finds 
\be
\partial^{++} \hat \phi^{++} = 0 \quad \Rightarrow \quad \hat \phi^{++} = 
\hat \phi^{ik}(x) w^{+}_iw^+_k~. \label{hatphi}
\ee

We observe that eq. \p{omeq} coincides with the pure 
harmonic part of the equation defining the Eguchi-Hanson metric 
in the harmonic superspace approach \cite{giot}. 
Its general solution was given in \cite{giot}, it depends on four 
arbitrary integration constants, that is, in our case, on four arbitrary 
functions of $x^{\mu i}$. However, these harmonic constants turn out to be 
unessential due to four hidden gauge symmetries of the set of equations 
\p{veq} - \p{xmineq}. One of them is the scale invariance 
$v{}' = v + \beta (x)$, while three remaining ones form 
an extra local $SU(2)$ symmetry \cite{iv2}. Using this gauge freedom  
one can gauge away four integration constants in $\omega $ and 
write a solution to eq. \p{omeq} in the following simple form  
\bea 
&& \omega = \sqrt{1 + R \hat\phi^2} \quad \Rightarrow \quad 
v = {1\over 2R}\; \mbox{ln} ( 1 + R \hat \phi^2 )~,  \label{vslut} \\
&& v^{++} = \partial^{++}v = {\hat \phi\; \hat \phi^{++} \over 
1 +R \hat \phi^2}~, \qquad \hat \phi \equiv
\hat \phi^{(ik)}(x)w^+_i w^-_k~. \label{vplplsol}
\eea 
One can restore the general form of the solution as it was 
given in \cite{giot}, acting on \p{vslut} by a finite form of 
the aforementioned hidden symmetry transformations. 
In \cite{iv2} we demonstrate that 
the whole effect of the full gauge 
$SU(2)$ transformation is reduced to the rotation of the $\tau $-world 
metric 
corresponding to the fixed-gauge solution \p{vplplsol} by some 
harmonic-independent non-singular matrix which becomes identity upon 
restriction to $x $- independent $SU(2)$ transformations. 
Thus in what follows we can stick to this solution.

Now we are prepared to solve eq. \p{xpluseq} (or \p{hatxeq}). This can be 
done in a full analogy with the hyper-K\"ahler TN case \cite{cmp},
based essentially upon the PG invariance \p{PG}.
Using \p{vplplsol}, we obtain 
\bea
&& \kappa^{++} = \partial^{++} \kappa~,  \nn 
&& (1)\;R > 0~, \ \kappa = {1\over \sqrt{R}} 
\;\mbox{arctan} \;\sqrt{R}\ \hat \phi~; \;\;  
(2)\;R < 0~, \ \kappa = {1\over \sqrt{|R|}} 
\;\mbox{arctanh}\; \sqrt{|R|}\ \hat \phi . \label{1kappa} 
\eea
For definiteness, in what follows we will choose the solution (1) in 
\p{1kappa}. 
Then, making the redefinition 
\be
\hat x^+ = \exp\{2i\kappa\} \tilde x^+~, \qquad 
\overline{\hat x^+} = \exp\{-2i\kappa\} \bar{\tilde x}^+~, \label{deftild}
\ee
we reduce \p{hatxeq} to 
\bea 
&& \partial^{++} \tilde x^+ = 0 \quad \Rightarrow \nn 
&& \tilde x^+ = x^iw^+_i~, \;  \bar{\tilde x}^+ = \bar x_i w^{+i} = 
- \bar x^iw^+_i~, \; \hat \phi = -2i\lambda x^{(i}\bar x^{k)} 
w^+_iw^-_k~, \label{solut}
\eea
where, in expressing $\hat \phi $, we essentially made use of the 
PG symmetry \p{PG}. 

Combining eqs. \p{defhat}, 
\p{vslut}, \p{deftild} and \p{solut} we can now write the expressions for 
$x^+$, $\bar x^+$ in the following form 
\be
x^+ = {1\over \sqrt{1 +R \hat \phi^2}} \exp\{2i\kappa\}\; x^iw^+_i~, 
\quad \bar x^+ = - {1\over \sqrt{1 +R \hat \phi^2}} 
\exp\{-2i\kappa\}\; \bar x^iw^+_i~, \label{solutx}
\ee
where $\kappa $ and $\hat \phi $ are expressed through $x^i, \bar x^i$ 
according to eqs. \p{1kappa}, \p{solut}. Comparing \p{solutx} with the 
general definition of the $x$ -bridges \p{xbridge}, we can identify 
$x^i, \bar x^i$ with the $\tau $- world coordinates, i.e. with the 
coordinates of the initial $4$-dimensional QK manifold. 

We still need to find $x^-, \bar x^-$ as functions of $x^i, \bar x^i$ 
and harmonics $w^{\pm}_i$ by solving eq. \p{xmineq} and its conjugate. 
Dropping intermediate technical steps (they involve a number of 
redefinitions), its general solution can be presented in the 
following form  
\bea
&& x^- = {1\over 2\lambda}\;{\sqrt{1 +R \hat \phi^2}\over (\lambda s) - 
i\hat \phi }
\left[\mbox{e}^{-2i\kappa(i\lambda s)} - \mbox{e}^
{2i\kappa(\hat \phi)} \right] \tilde x^-~, \nn 
&& \bar{x}^- = {1\over 2\lambda}\;{\sqrt{1 +R \hat \phi^2}\over (\lambda s) 
+ i\hat \phi}
\left[\mbox{e}^{-2i\kappa(i\lambda s)} - \mbox{e}^
{-2i\kappa(\hat \phi)} \right] \bar{\tilde x}^-~. \label{xminsol}
\eea
Here 
\be 
\tilde x^- = x^iw^-_i~, \quad \bar{\tilde x}^- = -\bar x^iw^-_i~,  \quad
s = x^i\bar x_i~, \quad
\kappa(i\lambda s) \equiv \kappa_0  = 
{i\lambda\over \sqrt{R}}\; \mbox{arctanh}\;\sqrt{R} (\lambda s)~.  
\label{defs}
\ee
For what follows it will be convenient to define 
\bea
A(s) \equiv 1 - R\lambda^2s^2~, \quad B(s) \equiv 1+4\lambda^2 s  + 
R \lambda^2 s^2~, \quad
C(s) \equiv 1+ Rs +R\lambda^2s^2~. \label{defABC}
\eea
Now we are ready to find explicit expressions for the two important 
quantities 
entering the general expression for the $\tau $-metric \p{12}, \p{13}:
\be 
1 - R \partial^{--} v^{++} = A \; 
{ 1 - R \hat \phi^2 \over ( 1 + R \hat \phi^2 )^2}~, \quad 
1-R(x\cdot H) = \frac{C}{A}\; \frac{1+ R \hat \phi^2}{1- R \hat \phi^2}~.
\label{exprpsi}
\ee 

As a next step towards the QK Taub-NUT metric, one needs to find 
the entries of the matrix $X^{+\mu i}_\nu \equiv \partial^+_\nu x^{\mu i}$  
by solving the set of algebraic equations 
\p{1}. In the complex notation, this set is divided into the two mutually 
conjugated ones, each consisting of four equations. It is clearly enough 
to consider one such set, e.g. 
\be \label{2}
X^{+ \rho k} \nabla_{\rho k} x^{-} = 1, \quad 
X^{+ \rho k} \nabla_{\rho k} x^{+} = 0, \quad
X^{+ \rho k} \nabla_{\rho k} \bar x^{\pm} = 0,
\ee
where $X^{+ \rho k} \equiv X^{+ \rho k}_1, 
(\bar X^{+ \rho k} \equiv - X^{+ \rho k}_2) $.
It is convenient to work with 
\be \label{3}
\hat X^{+ \rho k} = 
e^{R v} X^{+ \rho k}  = \sqrt{1 + R\hat \phi^2} \;X^{+ \rho k}~. 
\ee

It remains to calculate the transition matrix elements 
$\nabla_{\rho k}x^{\pm}, \nabla_{\rho k}\bar x^{\pm}$ entering 
eqs. \p{1}. This can be done straightforwardly, the corresponding 
expressions look rather involved and by this reason we 
do not quote them here explicitly (more details are given 
in \cite{iv2}). Surprisingly, the expressions for 
$\hat X^{+\rho k}$ prove to be much simpler: 
\bea 
\hat X^{+1k} &=& 
{1\over 4}\;[(3A +B)\epsilon^{kl} - 
4\lambda^2(A+C)x^{(k}\bar x^{l)}\;]\;w_l^+
e^{2i\kappa_0}\;, \nn
\hat{X}^{+2k} &=& \hat{(\partial^+ \bar x^k)} 
= \lambda^2\;(A+C)\;(\bar x^k \bar x^l) w^+_l e^{2i\kappa_0} 
\label{11}
\eea
(the remaining components can be obtained by conjugation).

It will be convenient to rewrite the metric \p{12}, \p{13} through 
$\hat X^{+\rho i}_\mu$
\bea 
g^{\rho i, \lambda k} &=& {1\over C \mbox{det}(\partial \hat{H})}  
\;\hat{G}^{\rho i, \lambda k}\;, \label{15} \\
\hat G^{\rho i, \lambda k} &=& 
(1+R\hat\phi^2)\; G^{\rho i, \lambda k} = 
\epsilon^{\omega \beta}\;[\;\partial^{--} 
\hat{X}^{+ \rho i}_\omega\;\hat{X}^{+ \lambda k}_\beta + 
(\rho i \leftrightarrow \lambda k) \;]\;. 
\label{16}
\eea

As the last step, one should compute $\mbox{det}(\partial \hat H)$. 
After some algebra, it can be represented in the following concise form 
\bea 
\mbox{det} (\partial \hat H) = -{1\over 2}\;{A\over C^3}(1- R\hat \phi^2) 
\;[\;
\epsilon^{\alpha \beta}\partial^{--}\hat{X}^{+\rho k}_\alpha 
\partial^{--}\hat{X}^{+\lambda l}_\beta \;][\;\nabla_{\rho k}x^{+\nu}
\nabla_{\lambda l}x^{+\mu}\epsilon_{\nu\mu}\;]~. \label{18} 
\eea
As a result of rather cumbersome, though straightforward computation 
one eventually gets the simple expression for $\mbox{det}(\partial \hat H)$
\be
\mbox{det} (\partial \hat H) = A^2\; {B\over C^3}\; e^{4i\kappa_0} = 
(1-R\lambda^2s^2)^2\; 
{1+2\lambda^2 s + \lambda^2 s(2+sR)\over (1+Rs + R\lambda^2s^2)^3}\; 
e^{4i\kappa_0}\;. 
\label{20}
\ee   
The harmonic dependence disappeared in $\mbox{det} 
(\partial \hat H)$, as it should be.

The calculation of this determinant is the most long part of the whole 
story. Once this has been done, the computation of the $\tau $ basis 
inverse metric amounts to the computation of entries of the matrix 
$\hat G^{\rho i, \lambda l}$. The final answer for the metric tensor 
is as follows 
\be\label{dist}
\left\{\begin{array}{lll} 
\displaystyle g_{1k, 1t}\ = & \displaystyle \frac{D}{C^2 B}\; 
(\bar x_k \bar x_t)\;, &\quad A=1-R\lambda^2s^2,  \\ [3mm]
\displaystyle g_{2k, 2t}\ = & 
\displaystyle \frac{D}{C^2 B}\; ( x_k x_t)\;, &\quad B=1+4\lambda^2s
+R\lambda^2s^2, \\ [3mm]
\displaystyle g_{1k, 2t}\ =& \displaystyle \frac 1{C^2 B}\;
[B^2 \epsilon_{kt} + D(\bar x_k x_t)\;]\;, &\quad C=1+Rs+R\lambda^2s^2. 
\end{array}\right. 
\ee
Here $D \equiv \lambda^2 (A+C)(A+B) = 2\lambda^2 (2+Rs)(1+2\lambda^2s)$. 
One should observe that this final expression is valid for any sign 
of the parameter $\,R\,$ even if, in the intermediate steps (see relation 
(\ref{1kappa}) for instance), the sign choice plays a significant role.

\vspace{0.4cm}    
\noindent {\bf 4.} To compare to the results in the literature \cite{egh} 
one has to use 
\cite{dev}
\be\label{sig1}
dx^i=x^i\left(\frac{ds}{2s}+i\frac{\sigma_3}{2}\right)-
\bar x^i\left(\frac{\sigma_2-i\sigma_1}{2}\right),\qquad
d\sigma_i=\frac 12\epsilon_{ijk}\,\sigma_j\wedge\, \sigma_k.
\ee
Using the notation $\ A\cdot B\equiv A^i\, B_i\ $ relation (\ref{sig1})
implies 
\be\label{sig2}
-\bar x\cdot dx=\frac{ds}{2}+
is\frac{\sigma_3}{2},\qquad dx\cdot d\bar x=\frac{ds^2}{4s}+
\frac s4(\sigma_1^2+\sigma_2^2+\sigma_3^2),\quad s=x\cdot\bar x.\ee
The metric given by (\ref{dist}) becomes
\be\label{dist2}
\frac 12\left[ \frac B{sC^2}\, ds^2+\frac{sB}{C^2}\,(\sigma_1^2+\sigma_2^2)
+\frac{sA^2}{C^2 B}\,\sigma_3^2\right].\ee

The most general Bianchi IX euclidean Einstein metrics can be deduced from 
Carter's results \cite{ca}. A convenient  standardization \cite{chv} is the 
following
\be\label{ca1}
d\tau^2=l^2\left \{\frac{r^2-1}{\Delta(r)}(dr)^2+4\  
\frac{\Delta(r)}{r^2-1} \ \sigma_3 ^2 
+(r^2-1) (\sigma_1 ^2 +\sigma_2 ^2)\right \},\ee
with
\be\label{ca2}
\Delta(r)=
\frac{-\Lambda l^2}{3}r^4+(1+2\Lambda l^2)r^2 -2\,M\,r+1+\Lambda l^2.\ee
These metrics are Einstein, with Einstein constant $\Lambda$ and 
isometry group $U(2)$. If we take $\, M=4/3\Lambda l^2+1\,$ the metric 
simplifies to
\be\label{ca3}
d\tau^2({\mathbb Q})=l^2\left \{\frac{r+1}{r-1}\frac{(dr)^2}{\Sigma(r)}
+4\  \frac{r-1}{r+1}\,\Sigma(r)\,\sigma_3 ^2
+(r^2-1) (\sigma_1 ^2 +\sigma_2 ^2)\right \},\ee
where now
\be\label{ca4}
\Sigma(r)=1-\frac{\Lambda\, l^2}{3}(r-1)(r+3).\ee
The identifications
\be\label{ca5}
\frac{r-1}{2}=(4\lambda^2-R)\frac s{1+Rs+R\lambda^2s^2},\qquad
\frac 43{\Lambda l^2}=\frac R{4\lambda^2-R},\ee
give the relation
\be\label{ca6}
4(4\lambda^2-R)\left[ \frac B{sC^2}\, ds^2
+\frac{sB}{C^2}\,(\sigma_1^2+\sigma_2^2)
+\frac{sA^2}{C^2 B}\,\sigma_3^2\right]=\frac{d\tau^2({\mathbb Q})}{l^2}.\ee
The quaternionic metric (\ref{ca3}) is complete for $\,\Lambda<0$ and is  
asymptotically Anti de Sitter. It has been considered recently  in \cite{hhp} 
under the name Taub-NUT-AdS metric and reveals itself a useful background for 
computing black-holes entropy.

\vspace{0.4cm}
\noindent {\bf 5.} In this paper we made the first practical use of 
the harmonic space formulation of the QK geometry \cite{gio} to 
compute a non-trivial QK metric, the four-dimensional quaternionic 
Taub-NUT metric. As we were convinced, the harmonic space techniques, 
like in the HK case \cite{{cmp},{HK},{giot}}, allows one 
to get the {\it explicit} form of the QK metric starting from a given 
QK potential and following a generic set of rules. It would be 
interesting to apply this approach to find the QK analogs of 
some other interesting 4- and higher-dimensional HK metrics, in particular, 
the quaternionic Eguchi-Hanson metric and the quaternionic 
generalization of the multicenter metrics of Gibbons and Hawking \cite{GH}. 

Finally, we note that the HK Taub-NUT metric plays an important role 
in the modern $p$-branes realm, yielding an essential part of one of 
the fundamental brane-like classical solutions of $D=11$ supergravity, the 
so-called `Kaluza-Klein monopole' (see, e.g. \cite{stelle}). It would be of 
interest to reveal possible brane implications of the QK Taub-NUT metric 
constructed here. 

\vspace{0.4cm}
\noindent{\bf Acknowledgement.} E.I. thanks the Directorate of 
LPTHE, Universit\'e Paris 7, 
for the hospitality extended to him during the course of this work. His 
work was partly supported by the grant of Russian Foundation of Basic 
Research RFBR 96-02-17634 and by INTAS grants INTAS-93-127-ext, 
INTAS-96-0538.

\end{document}